\documentclass[aps,prb,twocolumn,superscriptaddress,showpacs,preprintnumbers,longbibliography]{revtex4-1}
\usepackage[colorlinks,bookmarks=false,citecolor=blue,linkcolor=red,urlcolor=blue]{hyperref}
\usepackage{physics}
\input pdfcolor.tex
\usepackage{colortbl,amsthm,txfonts}
\usepackage{verbatim}
\usepackage{graphicx}
\usepackage{epsfig}
\usepackage{dcolumn}

\usepackage{bm}

\usepackage{amsmath,amssymb}

\makeatletter
\renewcommand*\env@matrix[1][*\c@MaxMatrixCols c]{%

\hskip -\arraycolsep
\let\@ifnextchar\new@ifnextchar
\array{#1}}
\makeatother
\usepackage{appendix}

\begin{document}

\title{Non Fermi liquid signatures across strain engineered metal-insulator transition in line-graph lattices} 
\author{Shashikant Singh Kunwar}
\affiliation{Department of Physics and Astronomy, University of Iowa, \\
Iowa City, Iowa 52242, United States of America}
\author{Madhuparna Karmakar}
\email{madhuparna.k@gmail.com}
\affiliation{Department of Physics and Nanotechnology, SRM Institute of Science and Technology,   \\ 
Kattankulathur, Chennai 603203, India}

\begin{abstract}
Controlling the properties and thus the functionalities of correlated electron systems via externally tunable perturbations 
has always remained a cherished goal in quantum condensed matter physics. Recently, straintronics has proved 
to be one such external control which can dictate the quantum phases and transitions in materials via the 
reconstruction of their electronic band structure. A particularly intriguing scenario arises in the context of flat band 
line-graph lattices wherein straintronics is found to bring forth non trivial phase transitions. This paper reports the 
phase transitions and thermal scales across the Lieb/Kagome interconversion in the electronic interaction-strain-temperature 
space. Based on the thermodynamic, spectroscopic and transport signatures across the strain tuned interconversion of 
these line-graph lattices we have mapped out the low temperature phases and thermal transition scales, numerically determined 
using non perturbative calculations. While at the low temperatures,  interaction-strain plane is spanned by magnetically correlated 
insulators, flat band induced weak transiently localized insulators and non Fermi liquid metallic phases, thermal fluctuations aid in to 
stabilize coexistent magnetic correlations. Apart from quantifying the magnetic transition scales in this system our results on the spectroscopic 
and transport signatures distill the strain tuned metal-insulator transition and crossover scales which exhibit variable transport scaling exponents. 
\end{abstract}

\date{\today}
\maketitle

\section{Introduction}
Quantum materials are inherently sensitive to the external perturbations bringing forth via band structure reconstruction; exotic phases, their competition and the scales associated therein. A particularly intriguing scenario arises in geometrically frustrated materials with multiple electronic bands and flat bands. An example being the Kagome metal, currently in vogue owing to its recent material realization and prospective artificial engineering \cite{higo_nature2015,parkin_sciadv2016,shin_natmat2017,otani_nature2019,behnia_natcom2019,hess_prb2019,wills_jpcm2009,zhang_advmat2017,checkelsky_nature2018,wang_nature2018,zhang_prl2018,wang_apl2019,tranquada_prl2019,kondo_prb2020,chen_natphys2018,lei_natcom2018,hasan_natphys2019,chen_science2019,shen_apl2019,analytis_natcom2020,madhavan_natcom2021,ochiai_prb2018,ochiai_jspj2019,checkelsky_apl2019,comin_natmat2019,zhang_prb2020,mcguire_prm2019,wilson_prm2021,wilson_prl2020,bredas_materhor2022,schiffrin_advfuncmat2021}. Interestingly, the Kagome lattice shares many of its characteristics with another line-graph lattice viz. the Lieb lattice \cite{liu_prb2019,denx_pra2023,pereira_prb2023,huang_natcom2020,liu_natcom2019,montambaux_prb2020}. 

Naturally, there has been a flurry of research to look for a potential tuning parameter or perturbation which not just aids in to smoothly traverse from one lattice to the other but is experimentally accessible. In this quest the perturbation which has repeatedly made its appearance in the literature is the applied strain, experimentally accessible and a clean substitute of chemical doping or disorder. Based on the non interacting model Hamiltonian it was shown that straintronics across the Lieb/Kagome interconversion leads to emergent Dirac cones via the deconstruction and the eventual reconstruction of the electronic band structure. The system undergoes topological phase transition across this interconversion which are quantified in terms of Chern number \cite{liu_prb2019,denx_pra2023,pereira_prb2023,huang_natcom2020,liu_natcom2019,montambaux_prb2020}. These noninteracting toy models are however, far removed from the practical Lieb or Kagome materials where electronic correlations dominate the physics and geometric frustration allows for novel quantum phases and phase competitions, as have been observed in the recently discovered Kagome materials such as, Mn$_{3}$Sn \cite{higo_nature2015,parkin_sciadv2016,shin_natmat2017,otani_nature2019,behnia_natcom2019,hess_prb2019}, Fe$_{3}$Sn$_{2}$ \cite{wills_jpcm2009,zhang_advmat2017,checkelsky_nature2018,wang_nature2018,zhang_prl2018,wang_apl2019,tranquada_prl2019,kondo_prb2020}, Co$_{3}$Sn$_{2}$S$_{2}$ \cite{chen_natphys2018,lei_natcom2018,hasan_natphys2019,chen_science2019,shen_apl2019,analytis_natcom2020,madhavan_natcom2021}, Gd$_{3}$Ru$_{4}$Al$_{12}$  \cite{ochiai_prb2018,ochiai_jspj2019}, FeSn \cite{checkelsky_apl2019,comin_natmat2019,zhang_prb2020,mcguire_prm2019} and the AV$_{3}$Sb$_{5}$ series (A= K, Rb, Cs) \cite{wilson_prm2021,wilson_prl2020}. 

A possible engineered set up to realize this lattice interconversion is the metal organic framework (MOF) which is a promising avenue owing 
to its broad tunability in terms of material composition, topology and choice of substrates, all of which contribute to their advance functionalities \cite{bredas_materhor2022,schiffrin_advfuncmat2021}. The Lieb and Kagome band structures serve as the fundamental structural motifs of two-dimensional (2D) MOF.  They have been experimentally realized in compounds such as, 
 9,10-dicyanoanthracene-copper (DCA-Cu) \cite{yan_advfuncmat2021,schiffrin_advfuncmat2021,yan_acsnano2021}, MCl$_{2}$(pyrazine)$_{2}$ \cite{clerac_natcom2022},  and 1,2,4,5-tetracyanobenzene (TCNB) \cite{schneider_acsnano2024}. 

Numerical investigation of the quantum phases in presence of strong electronic correlations across this interconversion is a formidable task. This 
is particularly true in the Kagome limit where strong geometric frustration along with the multiple electronic bands in these materials render the existing non perturbative numerical approaches redundant, owing to severe fermionic sign problem and system size restrictions \cite{tsunetsugu_jpcm2007,asano_prb2016,hatsugai_prl2019,ohashi_prl2006,maekawa_prl2005,janson_prb2021,paiva_prb2023,held_rmp2018,held_prb2021,toschi_prb2016,thomale_prr2024,zhu_prbl2021}. Recently we have demonstrated that such numerical constrains can be eased out 
to an extent based on the static path approximated (SPA) Monte Carlo technique which allows one to access the low temperature physics of these systems for a reasonably large system size \cite{shashi_kagome2024,shashi_reentrant2025}.  Based on the half filled Kagome Hubbard model (KHM) with the anisotropic nearest neighbor hopping amplitudes mimicking the applied strain,  it was shown that in the weak coupling regime  continuous tuning of the applied strain leads to a re-entrant metal-insulator transition and crossover.  The intermediate strain regime hosts 
a non Fermi liquid (NFL) metal cradled between a gapped and a gapless insulating phases \cite{shashi_reentrant2025}.

In this paper we take a step further in this direction and investigate the comprehensive three-dimensional interaction-strain-temperature 
($U-\eta-T$) parameter apace across the Lieb/Kagome interconversion. The system is modeled using the half filled KHM with anisotropic hopping, which we analyze in terms of our extensive results on the thermodynamic, spectroscopic and transport signatures of this system. Our primary inferences based on this work include: (i) applied strain in the interaction plane brings forth magnetically correlated insulators, NFL metal and 
flat band induced weakly localized insulating phases, (ii) the metal-insulator transition and crossover in this system are controlled by the applied strain with the variable scaling exponents of the electronic transport signatures attesting to the deviation from the Fermi liquid description, (iii) thermal fluctuations promote thermodynamic phases with competing and coexisting magnetic order via stabilizing short range sub-dominant magnetic correlations. 

The remaining of this paper is structured as follows: in Section 2 we discuss the model Hamiltonian and the numerical approach used to understand the transitions in the $U-\eta-T$ space, along with the relevant indicators to characterize the thermodynamic phases and phase transitions. Section 3 contain the results of this work presented in terms of the low temperature signatures while Section 4 discusses the finite temperature phases and the thermal scales. We conclude in Section 5 following a discussion of our observations and their potential applications.  

\section{Model and Method}
We model this multiband system using the 2D Hubbard Hamiltonian defined on the strain tuned Lieb/Kagome lattice \cite{janson_prb2021,paiva_prb2023,shashi_kagome2024}, 
\begin{eqnarray}
\hat H & = & -\sum_{\langle ij \rangle, \sigma} t_{ij}(\hat c_{i, \sigma}^{\dagger}\hat c_{j, \sigma} + h. c. ) -\eta\sum_{\langle ij \rangle, \sigma} (\hat c_{i, \sigma}^{\dagger}\hat c_{j, \sigma} + h. c.) \nonumber \\ && - \mu\sum_{i, \sigma} \hat n_{i\sigma} + U \sum_{i}\hat n_{i, \uparrow}\hat n_{i,\downarrow} \nonumber \\ && 
\end{eqnarray}

Here, $t_{ij}=t=1$ represents the hopping amplitude between nearest-neighbors and sets the reference energy scale for the system. The parameter $\eta$ characterizes the applied shear strain, allowing interpolation between the Lieb ($\eta=0$) and the Kagome ($\eta=1$) limits.  The on-site Hubbard interaction is given by $U>0$, and the chemical potential $\mu$ is tuned to ensure a half-filled lattice. To render the model computationally tractable, the interaction term is decoupled using the Hubbard-Stratonovich (HS) transformation \cite{hs1,hs2}, which introduces auxiliary bosonic fields: a vector field ${\bf m}_{i}(\tau)$ that couples to the spin and a scalar field $\phi_{i}(\tau)$ that couples to the charge \cite{shashi_kagome2024}. SPA is based on the adiabatic (slow boson) approximation wherein the bosonic fields are treated as classical, 
serving as a static, randomly fluctuating disordered background to the fast moving fermions \cite{ciuchi_scipost2021,fratini_prb2023,kivelson_pnas2023}. The adiabatic approximation allows us to access the real frequency dependent quantities without requiring an analytic continuation. This is particularly useful in accessing the low temperature physics of geometrically 
frustrated systems such as, Kagome materials.  The adiabatic approximation is valid for $T > T_{FL}$ where, $T_{FL}$ corresponds  to the 
Fermi liquid coherence temperature \cite{shashi_kagome2024,ciuchi_scipost2021,fratini_prb2023,kivelson_pnas2023}. 
\begin{figure}
\begin{center}
\includegraphics[height=8cm,width=8.5cm,angle=0]{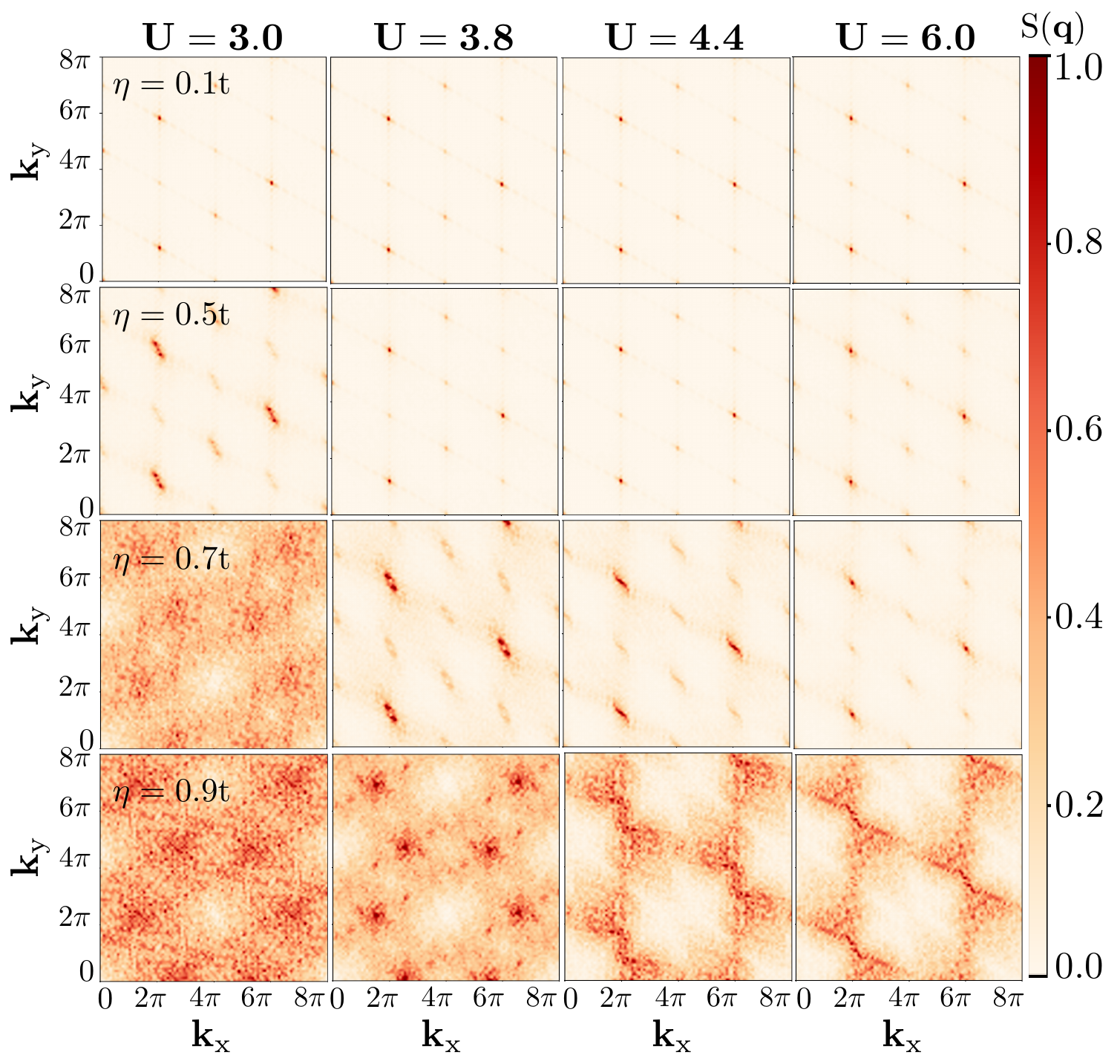}
\caption{Static magnetic structure factor ($S({\bf q})$) at $T=0.01t$ for selected $U-\eta$ cross sections.  At large strain, magnetic correlations 
set in for $U>3t$, the intermediate interaction $U=3.8t$ hosts a precursor Coulomb phase while the strong coupling regime ($U=4.4t$ and 
$U=6.0t$) is characterized by a Coulomb phase.}
\label{fig1}
\end{center}
\end{figure}

Within the framework of SPA, the $\phi_{i}$ field is treated at the saddle point level such that, $\phi_{i} \rightarrow \langle \phi_{i}\rangle = \langle n_{i}\rangle U/2$ (where $\langle n_{i}\rangle$ is the number density of the fermions), while the complete spatial fluctuations of the ${\bf m}_{i}$ field are retained. The resulting  effective Hamiltonian reads as \cite{shashi_kagome2024,shashi_reentrant2025}, 
\begin{eqnarray}
\hat H_{eff} & = & -t\sum_{\langle ij\rangle, \sigma}(\hat c_{i\sigma}^{\dagger}\hat c_{j\sigma}+h.c.) -\eta\sum_{\langle ij \rangle, \sigma} (\hat c_{i, \sigma}^{\dagger}\hat c_{j, \sigma} + h. c.) \nonumber \\ && -\tilde \mu \sum_{i\sigma} \hat n_{i\sigma}  - \frac{U}{2}\sum_{i}{\bf m}_{i}.{\bf \sigma}_{i}  + \frac{U}{4}\sum_{i}m_{i}^{2}
\end{eqnarray}
where,  the final term in $H_{eff}$ accounts for the stiffness cost of the classical bosonic field ${\bf m}_{i}$; ${\bf \sigma}_{i} = \sum_{a,b} \hat c_{ia}^{\dagger} \sigma_{ab} \hat c_{ib} = {\bf s}_i$ represents the fermions \cite{shashi_kagome2024,shashi_reentrant2025}.
The various phases are characterized based on the fermionic correlators: (i) static magnetic structure factor,  $S({\bf q}) = \frac{1}{N^{2}}\sum_{ij}\langle {\bf m}_{i}.{\bf m}_{j}\rangle e^{i {\bf q}.({\bf r}_{i}-{\bf r}_{j})}$, where ${\bf q}$ corresponds to the magnetic ordering wave vector and $N$ is the number of lattice sites. The notation $\langle ... \rangle$ denotes the Monte Carlo configurational average. (ii) Single particle density of states (DOS),  $N(\omega) = (1/N)\sum_{n}\langle \delta(\omega - \epsilon_{n})\rangle$ where, $\epsilon_{n}$ are the eigenvalues of a single equilibrium configuration. (iii) Optical conductivity,  $\sigma(\omega) =  \frac{\sigma_{0}}{N} \sum_{\alpha, \beta} \frac{f(\epsilon_{\alpha})-f(\epsilon_{\beta})}{\epsilon_{\beta}-\epsilon_{\alpha}} \vert \langle \alpha \vert J_{x} \vert \beta \rangle \vert^{2} \delta(\omega - (\epsilon_{\beta}-\epsilon_{\alpha}))$, 
where, the current operator $J_{x}$ is defined as, $J_{x}  =  -i\sum_{i, \sigma, \vec \delta} [{\vec \delta}t_{\vec \delta}c_{{\bf r}_{i}, \sigma}^{\dagger}c_{{\bf r}_{i}+\vec \delta, \sigma} - H. c.]$,  with $f(\epsilon_{\alpha})$ being the Fermi function and $\epsilon_{\alpha}$ and $\vert \alpha\rangle$ are respectively, the single particle eigenvalues and eigenvectors of $H_{eff}$ in a given background of $\{{\bf m}_{i}\}$. (iv) dc conductivity, $\sigma_{dc}$, which is the $\omega \rightarrow 0$ limit of $\sigma(\omega)$, with $\sigma_{0}=\frac{\pi e^{2}}{\hbar}$ in 2D. (v) Spectral function, $A({\bf k}, \omega) = -(1/\pi){\mathrm{Im}} G({\bf k}, \omega)$ where, $G({\bf k}, \omega) = lim_{\delta \rightarrow 0} G({\bf k}, i\omega_{n})\vert_{i\omega_{n} \rightarrow \omega + i\delta}$. $G({\bf k}, i\omega_{n})$ is the imaginary frequency transform of $\langle c_{\bf k}(\tau)c_{\bf k}(0)^{\dagger}\rangle$. The results presented in this paper corresponds to a system size of $3\times L^{2}$, with $L=18$ unless specified otherwise.  

\section{Low temperature regime}
We analyze our results by classifying them in terms of the temperature regimes.  We consider $T \sim 0.01t$, as the representative of 
the low temperature regime of the system. The thermal regime ($T > 0.01t$) is analyzed across different temperature cross sections. 
Note that Mermin-Wagner theorem restricts long range ordering in 2D for $T>0$ and the thermodynamic phases that we discuss in 
this paper are in fact (quasi) long range ordered. Further, the thermal phase transitions discussed herein are essentially Berezinskii-Kosterlitz-Thouless (BKT) transitions \cite{mermin_wagner}. The associated vortex-antivortex formations are however not observable in a 
finite sized system. 
\begin{figure}
\begin{center}
\includegraphics[height=6.5cm,width=7.0cm,angle=0]{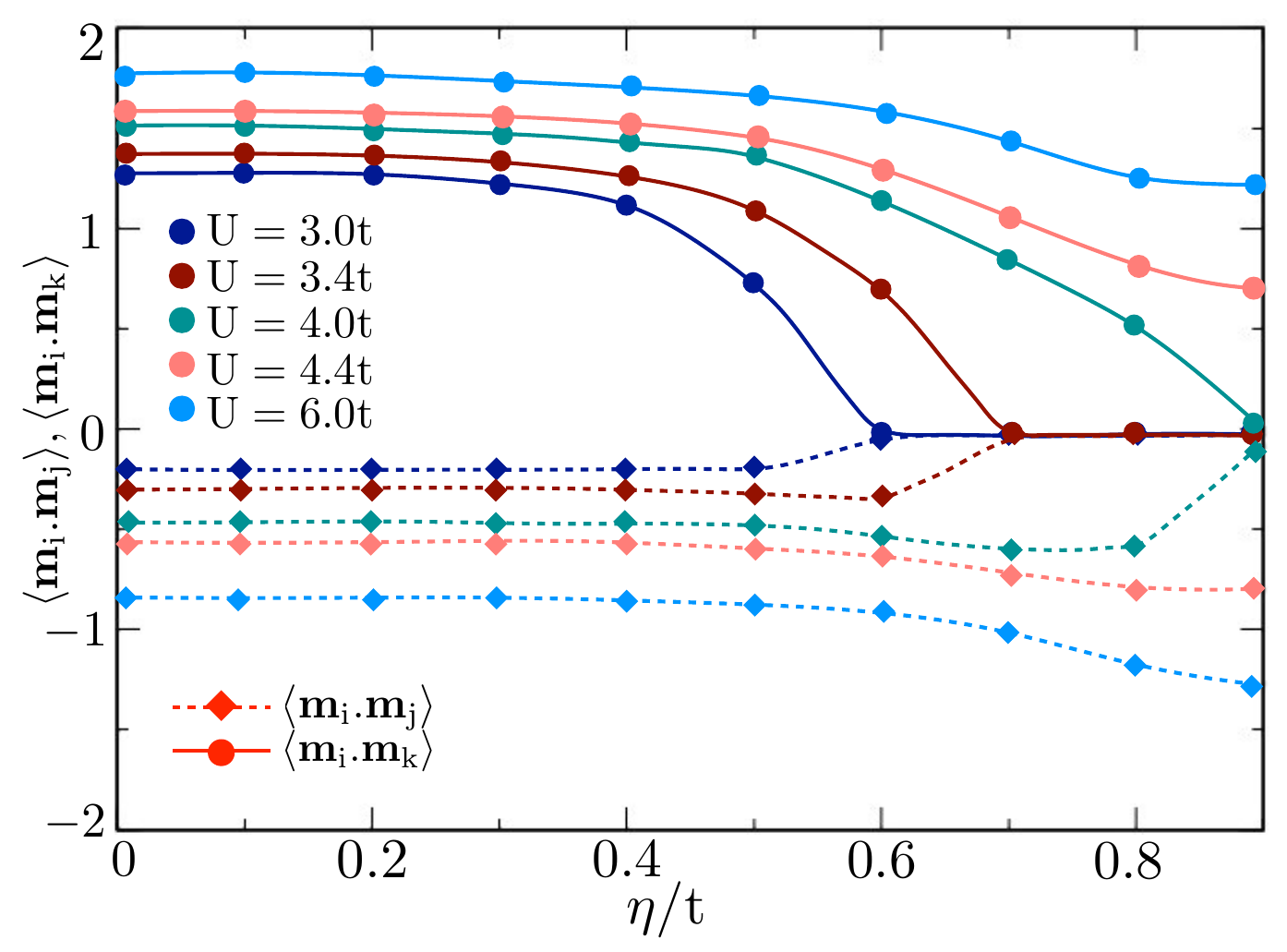}
\caption{Strain dependence of the NN ($\langle {\bf m}_{i} .{\bf m}_{j}\rangle$) and the NNN ($\langle {\bf m}_{i}.{\bf m}_{k}\rangle$) correlations 
at selected interactions, at $T=0.01t$. A positive sign of the magnitude indicates the FM correlation while the AF correlation is characterized by 
the negative sign.}
\label{fig2}
\end{center}
\end{figure}

\textit{Magnetic correlations and thermodynamic phases:}
We begin the discussion of the low temperature phases with the magnetic correlations across the $U-\eta$ plane. The static 
magnetic structure factor $S({\bf q})$ is shown in Fig.\ref{fig1}. The real space correlation function 
$\langle {\bf m}_{i} .{\bf m}_{j}\rangle$, corresponding to the nearest neighbor (NN) antiferromagnetic (AF) and 
$\langle {\bf m}_{i}.{\bf m}_{k}\rangle$ corresponding to the next nearest neighbor (NNN) ferromagnetic (FM) correlations, are 
shown in Fig.\ref{fig2}.  Across the range of interactions the low strain regime hosts a (quasi) long range magnetic order, dominated 
by the FM correlations, with subdominant contribution from the AF correlations. Quasi long range AF correlations begin to dominate for 
$\eta \gtrsim 0.5t$ depending on the interaction strength ($U/t$), giving rise to a $\sqrt{3}\times \sqrt{3}$ ordered precursor Coulomb 
phase \cite{shashi_kagome2024} at intermediate coupling. A Coulomb phase is realized close to the Kagome limit of $\eta=t$ at strong 
coupling, quantified in terms of the $\sqrt{3}\times \sqrt{3}$ order along with the pinch points in the $S({\bf q})$ maps, which are 
salient to geometric frustrations  \cite{udagawa_prb2018,zhitomirsky_prb2008,shanon_prb2018,iqbal_prr2023,gingras_science2001,bramwell_science2009,zaharko_prb2018}. 
\begin{figure}
\begin{center}
\includegraphics[height=9cm,width=8.5cm,angle=0]{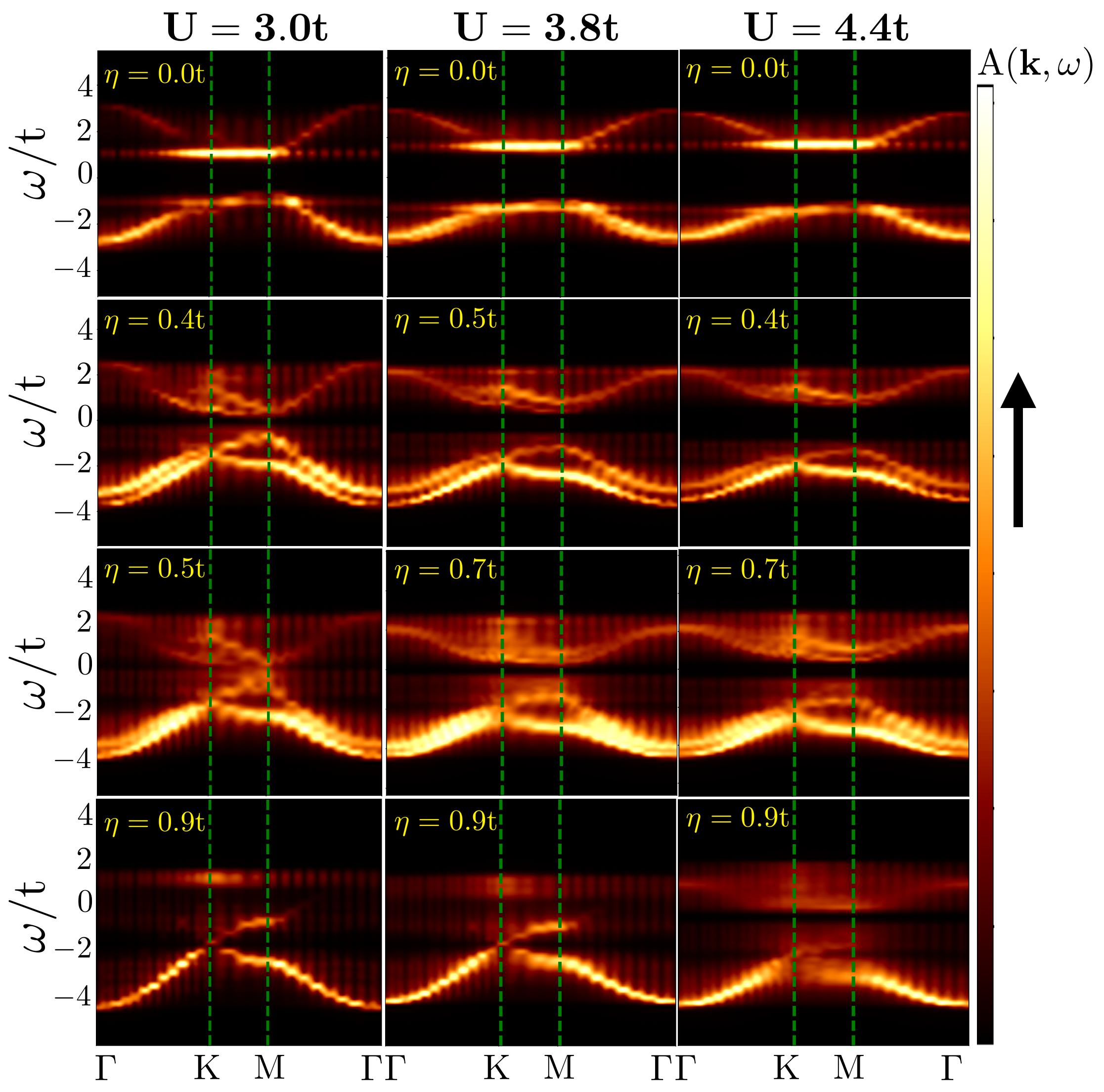}
\caption{Spectral function ($A({\bf k}, \omega)$) maps at $T=0.01t$ and selected $U-\eta$ cross sections showing the band structure reconstruction across the Lieb/Kagome interconversion.}
\label{fig3}
\end{center}
\end{figure}

\textit{Electronic band structure:} Fig.\ref{fig3} shows the spectral function ($A({\bf k},\omega)$) maps at selected $U-\eta$ cross sections, at $T=0.01t$. The non interacting Lieb lattice ($\eta=0.0$) is characterized by a flat band at the Fermi level and two dispersive bands which touches the flat band at the $M$-point \cite{liu_natcom2019,lieb_prl1989}. Interaction, as observed from Fig.\ref{fig3},  opens up an insulating Mott gap at the Fermi level, with the underlying magnetic phase being dominated by FM correlations. We demarcate this phase as the ferromagnetic insulator (FM-I). The dispersive bands, broadened due to interaction now overlaps the flat band along the $K-M$ trajectory. 

Applied strain has three-fold effect on the electronic band structure: (i) the overall spectrum shifts in energy, (ii) the dispersive bands undergo reconstruction, gives rise to emergent van Hove singularities and opens up topological gaps w. r. t. the flat band and (iii) a gapless spectra is realized at intermediate interactions and strain, indicating the collapse of the insulating (FM-I) phase. At strong interactions, the single particle spectrum is largely gapped across the $\eta/t$ regime.  
 
Across the range of interactions the large strain regime of $\eta=0.9t$ hosts a dominant high energy flat band, akin to the Kagome band structure. Interaction shifts this high energy flat band closer to the Fermi level as compared to the non interacting Kagome lattice where the flat band is 
at $\omega \sim 2t$. This interaction controlled proximity of the flat band to the Fermi level brings forth intriguing transport signatures quantifying flat band induced dynamic localization of the single particle states \cite{checkelsky_natphys2024,si_natcom2024,shashi_kagome2024,shashi_reentrant2025}.

\begin{figure}
\begin{center}
\includegraphics[height=8.0cm,width=8.5cm,angle=0]{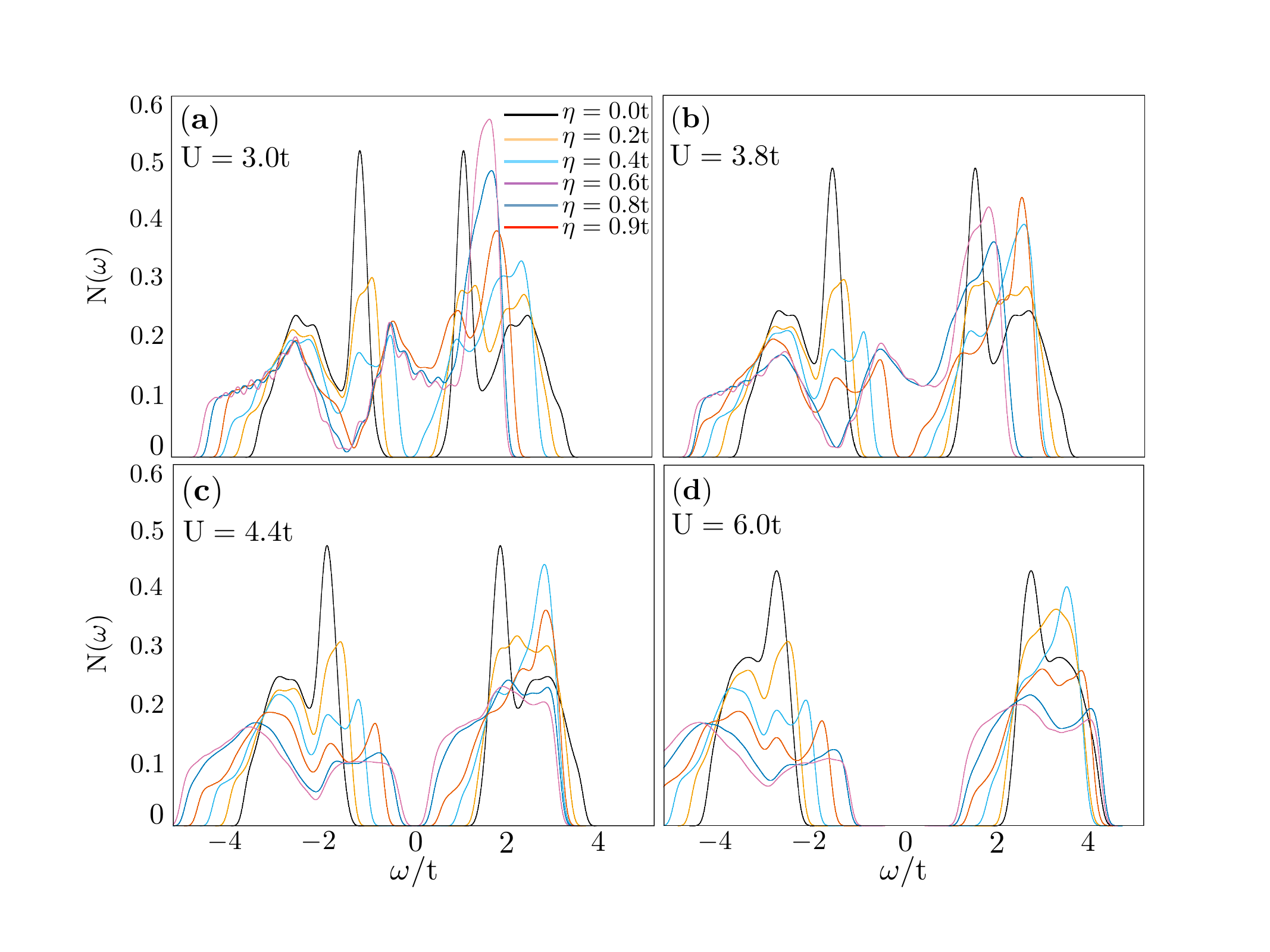}
\caption{Single particle DOS ($N(\omega)$) as a function of $\eta$ at selected interactions representative of: (a) weak ($U=3.0t$), 
(b-c) intermediate ($U=3.8t$, $U=4.4t$) and (d) strong ($U=6.0t$) coupling regimes, highlighting the strain controlled gap-gapless transition.}
\label{fig4}
\end{center}
\end{figure}

\textit{Single particle DOS:} Strain tuning of the low temperature single particle DOS at selected interactions is shown in Fig.\ref{fig4}. 
The FM-I in the weak strain regime is characterized by a robust spectral gap at the Fermi level.  Applied strain leads to the progressive suppression this gap such that, in the weak coupling regime a gapless, magnetically disordered state obtains corresponding to a paramagnetic metal (PM-M) (Fig.\ref{fig4}(a)). At intermediate coupling (Fig.\ref{fig4}(b)) the single particle DOS undergoes pronounced and progressive depletion of the spectral weight at $\omega \approx -1t$ and a gapless spectra at the Fermi level, corresponding to a ferromagnetic metal (FM-M) or an antiferromagnetic metal (AF-M), depending on the choice of $\eta/t$. The single particle spectra at the strong coupling, represented by $U=4.4t$ (Fig.\ref{fig4}(c)) and $U=6.0t$ (Fig.\ref{fig4}(d)) is gapped across the applied strain with the FM-I at weak and intermediate strain 
giving way to an AF Mott insulator (AF-MI) at large strain, accompanied by an underlying Coulomb phase.  

\textit{Optical transport:} The breakdown of the Fermi liquid description of strongly correlated quantum materials is quantified based on the 
optical absorption spectra. Measurement of low frequency behavior of the optical conductivity ($\sigma(\omega)$) provides unambiguous 
signatures of the NFL physics in various classes of materials with strong electronic correlations. Prominent examples include,  organic molecular charge transfer salt $\kappa$-[(BEDT-STF)$_{x}$(BEDT-TTF)$_{1-x}$]$_{2}$Cu$_{2}$(CN)$_{3}$ \cite{fratini_natcom2021}, organic Mott insulator $\kappa$-(ET)$_{2}$Cu[N(CN)$_{2}$]Cl \cite{kanoda_prl2020}, 
2D transition metal dichalcogenide (TMD)1T-TaS$_{2}$ \cite{trivedi_prl2014}, TMD heterostructure such as, MoTe$_{2}$/WSe$_{2}$ Moire superlattices \cite{mak_nature2021}, twisted bilayer WSe$_{2}$ \cite{pasupathy_nature2021} and more recently Kagome metals such as, Ni$_{3}$In \cite{checkelsky_natphys2024}, CsV$_{3}$Sb$_{3}$ \cite{tsirlin_prb2021} etc.  Recent results on the low frequency behavior of 
$\sigma(\omega)$ in the large strain regime at $U \sim t$ revealed a transiently localized insulating phase with {\it variable} optical conductivity scaling exponent ($\sigma(\omega) \propto 1/\omega^{\gamma}$) \cite{shashi_reentrant2025}. This gapless phase fails to establish any quasi long range magnetic correlation and is typified as a paramagnetic flat band induced insulator (PM-FI). The PM-FI  phase is weakly localized and is characterized by a small but finite Inverse Participation Ratio at the Fermi level,  $IPR(\omega=0)$,  in the thermodynamic limit. It was argued that the local moments originating from the flat electronic bands aid in to localize the fermions at and close to the Fermi level, hence the name \cite{shashi_kagome2024,shashi_reentrant2025}. The inferences drawn based on the numerical calculations align well with the observations  made based on the electrical transport measurements on Kagome metal Ni$_{3}$In, exhibiting pronounced NFL signatures \cite{checkelsky_natphys2024}. 
\begin{figure}
\begin{center}
\includegraphics[height=8.5cm,width=8.5cm,angle=0]{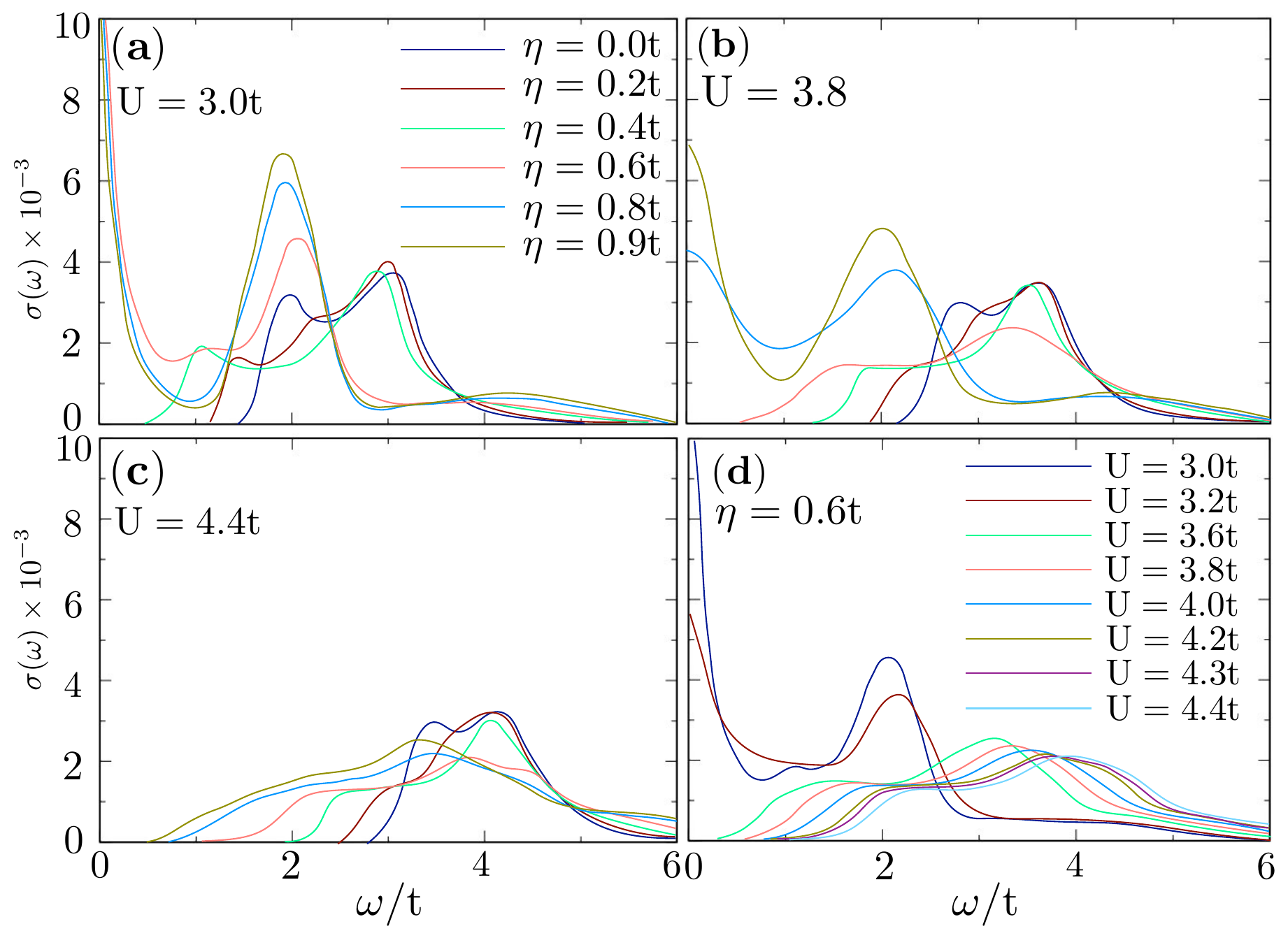}
\caption{(a)-(c): Strain dependence of the optical conductivity ($\sigma(\omega)$) at selected interactions, showing the NFL signatures 
in terms of the DDP and the intermediate frequency polaronic peak. (d) Variation of $\sigma(\omega)$ with interaction at the selected 
strain of $\eta = 0.6t$.}
\label{fig5}
\end{center}
\end{figure}

The deviation from the conventional $\sigma(\omega) \propto 1/\omega^{2}$ dependence results in a finite frequency displaced Drude peak (DDP), a salient feature of the NFL physics \cite{karlsson_scipost2017,grilli_prl2002,han_rmp2003,takagi_philmag2004,georges_prl2013,georges_prb2013,devereaux_science2019,dobro_prl2015,ciuchi_scipost2021,fratini_prb2023,kivelson_pnas2023}. Evidence of the NFL phase can further be observed from the intermediate frequency polaronic peak in $\sigma(\omega)$, arising out of the strong correlation between the fermions and the bosonic fields \cite{ciuchi_scipost2021,ciuchi_arxiv2024,kivelson_pnas2023,fratini_prb2023}. 

Fig. \ref{fig5}(a-c) shows the strain dependence of $\sigma(\omega)$ at selected interactions. The optical spectra in the FM-I phase is trivially gapped irrespective of the choice of $U/t$. However, in the regime of large $\eta$ there is finite low frequency spectral weight, in agreement 
with our observations on the spectral function and the single particle DOS.  At $U=3.0t$ (Fig.\ref{fig5}(a)) and $3.8t$ (Fig.\ref{fig5}(b)), for $\eta \gtrsim 0.5t$ and $\eta \gtrsim 0.7t$, respectively, $\sigma(\omega) \propto 1/\omega^{\gamma}$ as $\omega \rightarrow 0$  and $\gamma \neq 2$. The optical spectra is gapped across the strain tuned transition between the FM-I and AF-MI phases,  at $U=4.4t$ (Fig.\ref{fig5}(c)). Fig.\ref{fig5}(d) shows the interaction dependence of $\sigma(\omega)$ at a representative strain of $\eta=0.6t$. Based on the variable scaling exponent ($\gamma$) of the optical conductivity a transition from a gapless NFL metal at weak coupling to a gapped insulator at strong coupling could be ascertained. 
\begin{figure}
\begin{center}
\includegraphics[height=8.0cm,width=8.2cm,angle=0]{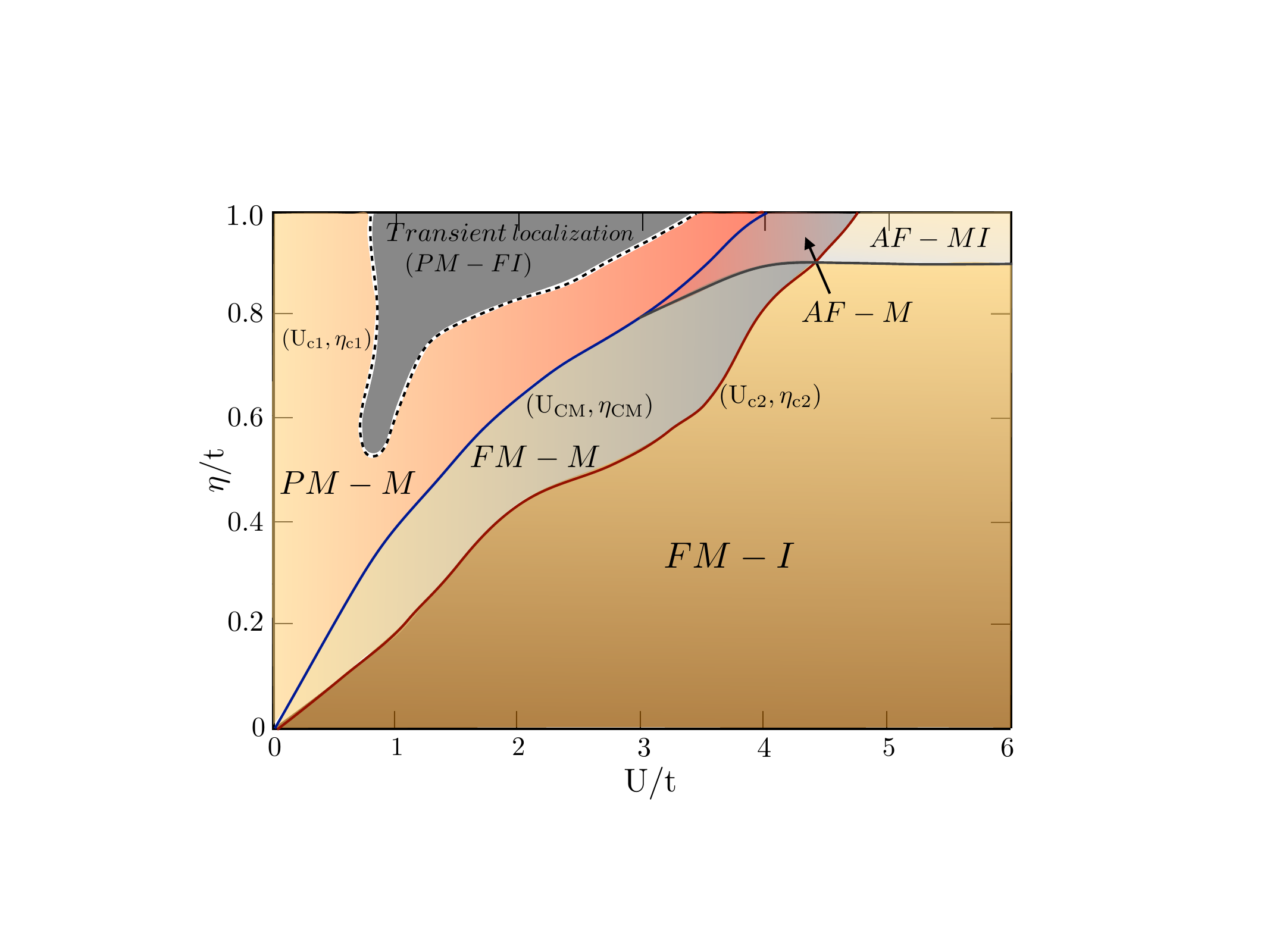}
\caption{Phase diagram in the $U-\eta$ plane,  at $T=0.01t$ showing the thermodynamic phases: (i )ferromagnetic insulator (FM-I), (ii) ferromagnetic metal (FM-M), (iii) paramagnetic metal (PM-M), (iv) paramagnetic (weak) transiently localized flat band insulator (PM-FI), (v) antiferromagnetic metal (AF-M) and (vi) antiferromagnetic Mott insulator (AF-MI). The critical points ($U_{c1}, \eta_{c1}$) quantifies the crossover between the PM-M and PM-FI phases while ($U_{c2}, \eta_{c2}$) demarcates the metal-insulator transition between the FM-I and FM-M phases as well as between the AF-M and AF-MI phases. The onset of the magnetic correlations is typified by the critical point ($U_{CM}, \eta_{CM}$). 
The dashed line shows interpolated boundary of the (weak) transiently localized phase.}
\label{fig6}
\end{center}
\end{figure}

\textit{Phase diagram:} Based on the aforementioned signatures we now map out the low temperature $U-\eta$ phase diagram,  at $T=0.01t$, in Fig.\ref{fig6}. The thermodynamic phases are demarcated based on the magnetic correlations and the single particle DOS. The metal-insulator transition and/or crossover scales are determined based on the low temperature behavior of the longitudinal resistivity ($\rho_{xx}$) (discussed in the next section). Fig.\ref{fig6} shows that the weak $\eta/t$ regime is FM-I across the range of 
interactions, dominated by NNN FM correlations with strongly suppressed single particle transport. The FM-I encompasses a large part of 
the $U-\eta$ plane giving way to a NFL metal at intermediate $U-\eta$ and to a dominantly AF phase at large $U-\eta$. The metallic phase 
at intermediate $U-\eta$ can further be classified into the ferromagnetically correlated metal (FM-M) and the disordered paramagnetic metal 
(PM-M).  These phases are by definition gapless and are NFL with pronounced deviation from the Fermi liquid description.  

The thermodynamic phase boundaries are associated with critical interactions and strains, ($U_{c1}, \eta_{c1}$), ($U_{CM}, \eta_{CM}$) and 
($U_{c2}, \eta_{c2}$) quantifying: (i) a metal-insulator crossover between the weak transiently localized PM-FI and PM-M at ($U_{c1}, \eta_{c1}$), across which no quantum symmetry breaking takes place and the system undergoes a gradual suppression of the single particle charge 
transport, (ii) a magnetic order-disorder transition across ($U_{CM}$, $\eta_{CM}$) between the PM-M/FM-M and PM-M/AF-M phases at the intermediate and large $U/t, \eta/t$, respectively, and (iii) a metal-insulator transition between the FM-I/FM-M at intermediate $U,/t \eta/t$ and between AF-MI/AF-M at large $U/t, \eta/t$, respectively, across ($U_{c2}$, $\eta_{c2}$). Finally, an insulator-insulator magnetic transition takes place at the large $\eta/t$ deep in the strong coupling regime between the FM-I and the AF-MI phases. 
\begin{figure}
\begin{center}
\includegraphics[height=8.5cm,width=8.5cm,angle=0]{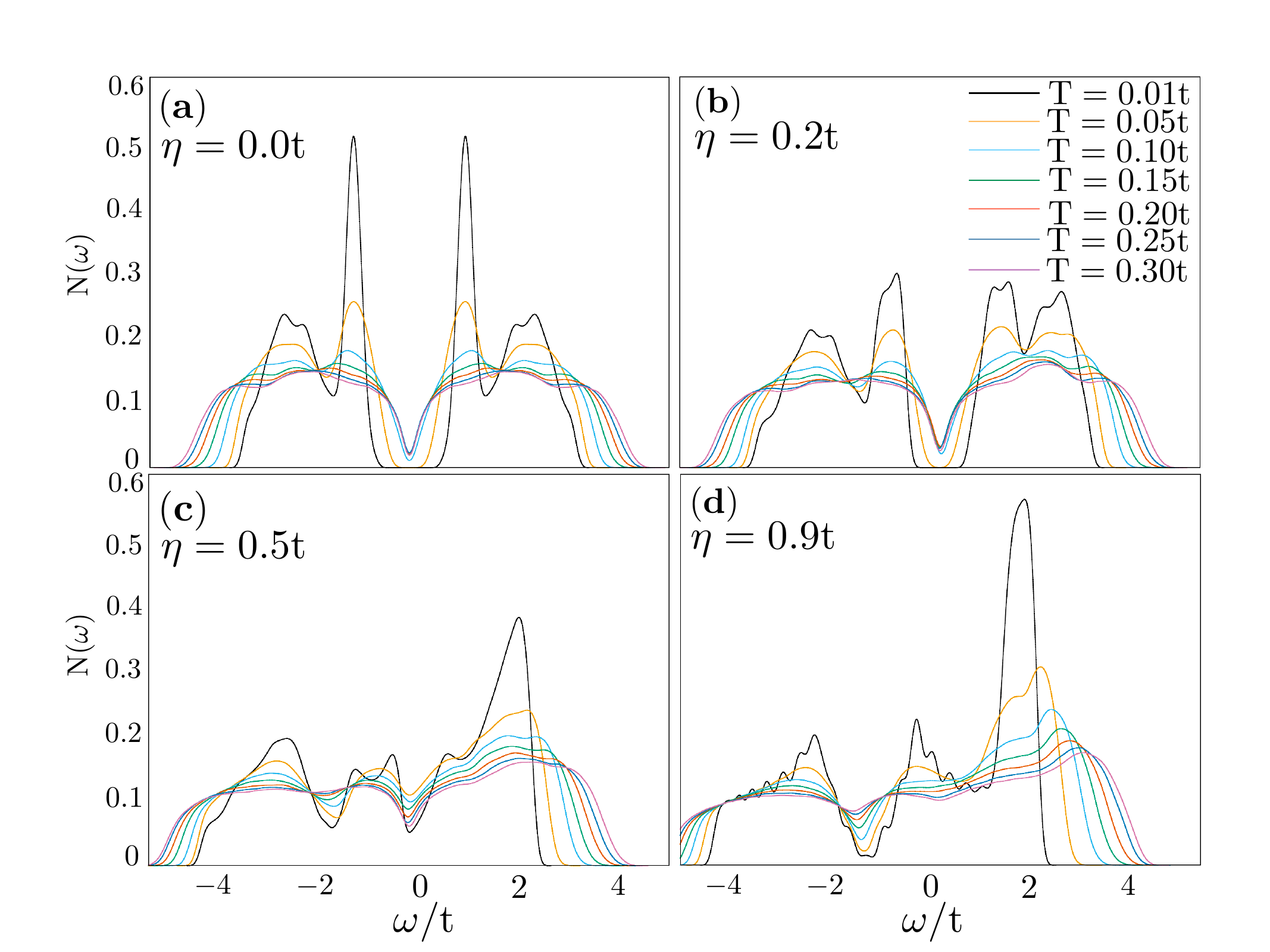}
\caption{Thermal evolution of the single particle DOS at $U=3.0t$ showing the thermal fluctuations induced accumulation of spectral weight 
and the origin of gapless phase for selected: (a) $\eta=0.0t$ (b) $\eta=0.2t$, (c) $\eta=0.5t$ and (d) $\eta=0.9t$.}
\label{fig7}
\end{center}
\end{figure}

\section{Thermal phases and scales}
With the low temperature phases in the $U-\eta$ plane in place we now select particular cross sections representative of the various 
phases of Fig.\ref{fig6},  to highlight the impact of thermal fluctuations on the thermodynamic phases and the associated thermal scales. 
Our analyses is based once again on the spectroscopic, transport and thermodynamic signatures of the system. 

\textit{Single particle DOS:}
Fig.\ref{fig7}(a-d) shows the thermal evolution of the single particle DOS at $U=3.0t$ for the selected $\eta=0.0t$ (Fig.\ref{fig7}(a)), 
$\eta= 0.2t$ (Fig.\ref{fig7}(b), $\eta= 0.5t$ (Fig.\ref{fig7}(c) and $\eta=0.9t$ (Fig.\ref{fig7}(d). At and close to the Lieb limit the (quasi) long 
range magnetic order is signified by the sharp van Hove singularities at the gap edges (Fig.\ref{fig7}(a-b)). Thermal fluctuations induced randomness of the local moments destabilize the (quasi) long range order and gives way to short range magnetic correlations. The 
corresponding single particle DOS is rendered gapless via the accumulation of spectral weight at the Fermi level that leads to the eventual 
closure of the gap.  Apart from the large transfer of spectral weights away from the Fermi level the single particle DOS is largely immune to temperature. 

The NFL metal at the intermediate strain (Fig. \ref{fig7}(c)), is essentially gapless with large spectral weight at the Fermi level at all 
temperatures. The broad high energy peak at $\omega \sim 2t$ signifies the strain tuned band structure reconstruction as the system moves 
away from the Lieb and towards the Kagome limit. For $T \gtrsim 0.05t$ the spectral weight begins to deplete giving rise to a dip at the Fermi 
level, an observation akin to the pseudogap formation in superconductors and correlated metals at finite temperatures. This suggests that even though the magnetic local moments fail to establish a (quasi) long range order, their short range correlations dominate the finite temperature regime at this $U-\eta$ cross section. 
\begin{figure}
\begin{center}
\includegraphics[height=8.5cm,width=8.5cm,angle=0]{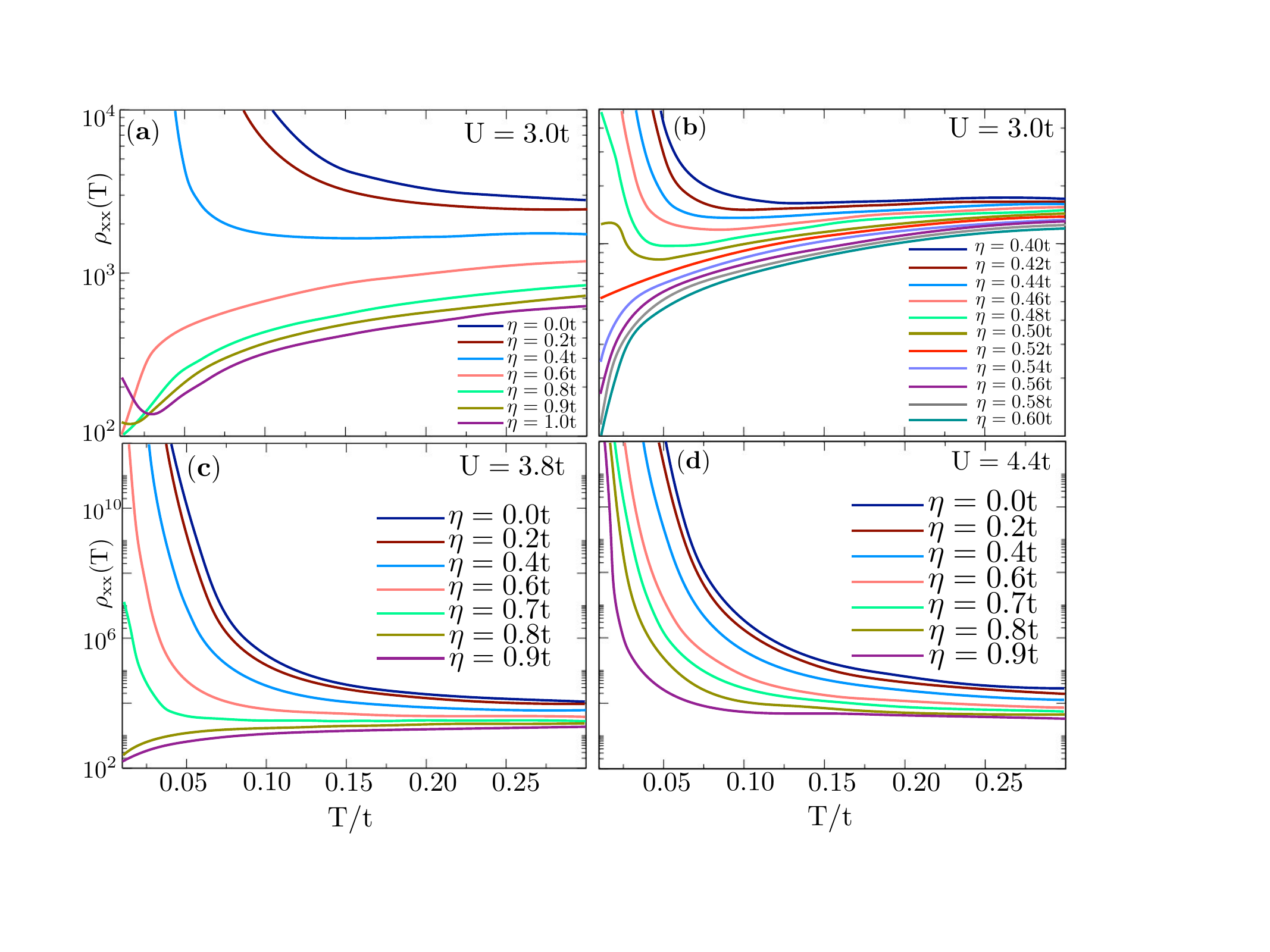}
\caption{(a-d) Temperature dependence of the longitudinal resistivity $\rho_{xx}(T)$ at selected $U-\eta$ cross sections demarcating the metal 
and the insulator phases based on the sign change of $d\rho_{xx}/dT$, as $T \rightarrow 0$. A $d\rho_{xx}/dT > 0$ signifies a metallic phase 
while an insulating phase is typified by $d\rho_{xx}/dT < 0$.}
\label{fig8}
\end{center}
\end{figure}

In the large strain regime (Fig. \ref{fig7}(d)), the single particle DOS at the Fermi level is largely featureless with a pronounced high energy 
peak at $\omega \sim 2t$ reminiscent of the Kagome limit. Thermal fluctuations progressively suppress this high energy peak as the 
electronic band structure is destroyed via thermal broadening. The single particle DOS shown in Fig. \ref{fig7}(a-d) are generic for the larger 
part of the $U-\eta$ plane. The strong coupling large strain AF-MI phase is quantified by a robust spectral gap at the Fermi level which 
remains largely immune to temperature. The weak transiently localized PM-FI phase is essentially gapless across the temperature regime \cite{shashi_reentrant2025}. 

\textit{Longitudinal resistivity:} One of the primary indicators of the Fermi liquid description of a metal and the deviations thereof is the low temperature behavior of its longitudinal resistivity ($\rho_{xx}$). For a conventional metal obeying the Fermi liquid description the electrical conductivity scales as the quasiparticle lifetime $\tau$ and the corresponding resistivity behaves as,  $\rho_{xx} \propto T^{2}$. This relation is often found to be violated in case of materials with strong electronic correlations wherein the Fermi surface is redundant and the quasiparticle lifetime is cut short \cite{fratini_natcom2021,kanoda_prl2020,mak_nature2021,pasupathy_nature2021,tsirlin_prb2021,pustogow_prb2023,ciuchi_scipost2021,devereaux_science2019}. In Fig.\ref{fig8} we present $\rho_{xx}(T)$ as a function of the applied strain $\eta/t$ at selected interactions. The distinction between the metallic and insulating phases are made based on the change in the sign of $d\rho_{xx}/dT$ as $T \rightarrow 0$,  such that, $d\rho_{xx}/dT > 0$ signifies a metallic phase while an insulating phase is typified by $d\rho_{xx}/dT < 0$.  At $U=3.0t$, (Fig.\ref{fig8}(a)-(b)) across the Lieb/Kagome interconversion the metal-insulator transition between the FM-I and NFL metal takes place at  $\eta \approx 0.52t$. 
In the large strain regime $\eta \gtrsim 0.8t$,  the observed $d\rho_{xx}/dT < 0$ indicates the re-entrant crossover of the metal to the weakly localized PM-FI phase.  
\begin{figure}
\begin{center}
\includegraphics[height=8.5cm,width=8.5cm,angle=0]{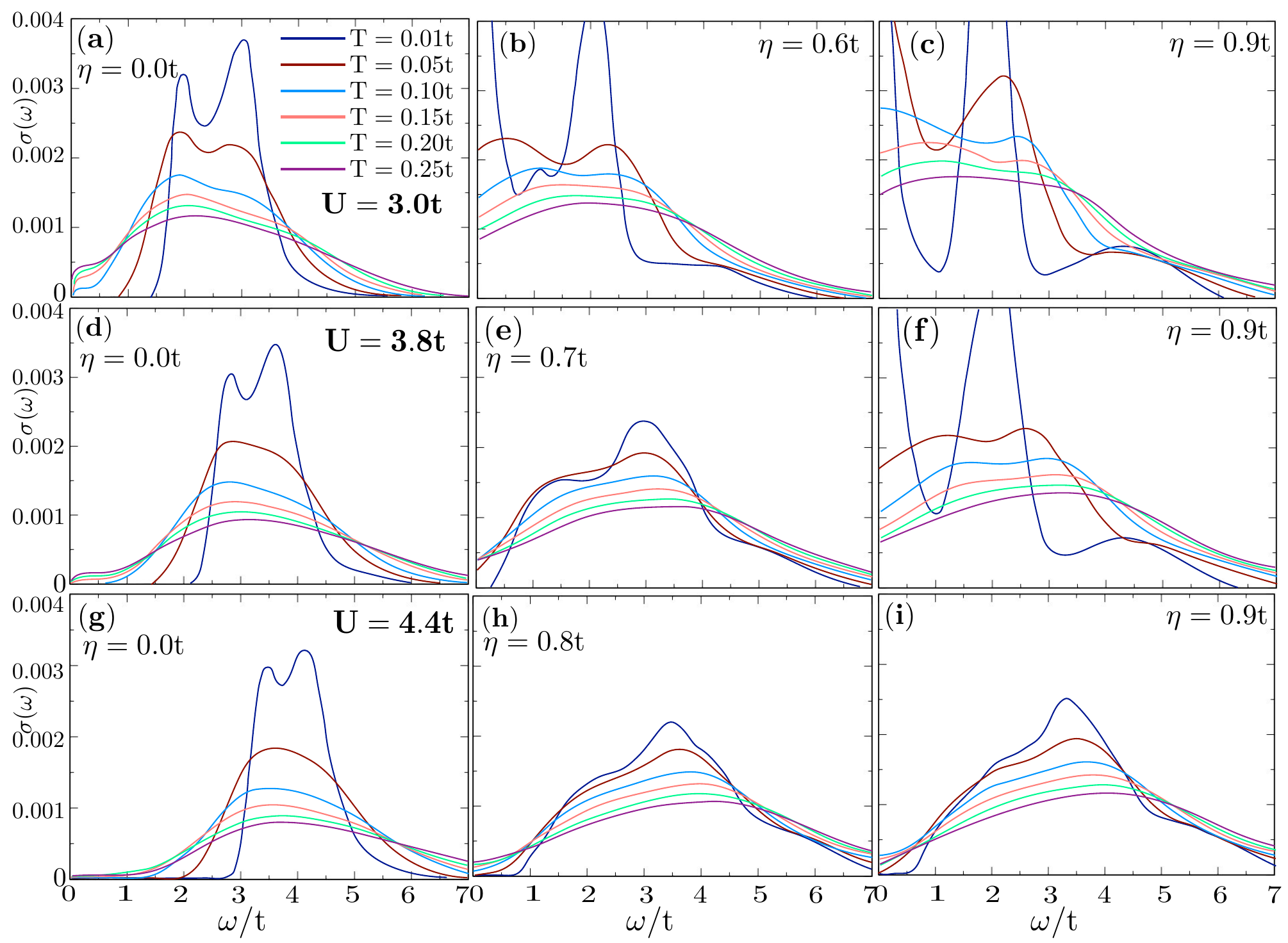}
\caption{(a-i): Thermal evolution of $\sigma(\omega)$ at selected $U-\eta$ cross sections showing the NFL metal and its thermal fluctuations induced crossover to the bad metal phase via the loss of the DDP and the polaronic peak.}
\label{fig9}
\end{center}
\end{figure}

At $U=3.8t$ the metal-insulator transition between the FM-I and NFL metal (FM-M and AF-M) takes place at $\eta \gtrsim 0.8t$, as shown in Fig.\ref{fig8}(c). Finally, Fig.\ref{fig8}(d) corresponds to $U=4.4t$ representative of a regime which lacks any metal-insulator transition. The system is 
a FM-I at weak and intermediate strain ($0 < \eta \lesssim 0.7t$) and undergoes transition to a AF-MI phase in the large strain regime (Fig.\ref{fig8}(d)). 

Across the $U-\eta$ plane, $\rho_{xx}(T)$ as $T \rightarrow 0$ exhibits pronounced deviation from the $\rho_{xx} \propto T^{2}$ behavior, 
attesting that the underlying phase is a NFL \cite{fratini_natcom2021,kanoda_prl2020,mak_nature2021,pasupathy_nature2021,tsirlin_prb2021,pustogow_prb2023,ciuchi_scipost2021,devereaux_science2019}. In the weak coupling regime ($U=t$) a $\eta$-dependent variable resistivity scaling exponent $\rho_{xx} = \rho(0)+AT^{\alpha}$, was recently reported, such that, at large strain $\alpha \sim 1.3, 1.5, 1.9$ for $\eta = 0.75, 0.85, 0.95$, respectively \cite{shashi_reentrant2025}. On the other hand the metallic regime at the intermediate strain $0.4 \lesssim \eta < 0.7$ for $U=t$ exhibits a {\it linear}-T ($\alpha=1$) dependence of $\rho_{xx}(T)$.  The variable resistivity scaling exponent is considered to be a hallmark of NFL physics.  Such variable scaling exponents have been experimentally observed in heavy fermion materials \cite{assmus_prl1998,steglich_prl2000,lonzarich_jpcm1996,ramazashvili_jpcm2001}, disorder-controlled metal-Mott insulator transitions \cite{dagotto_prl2017}, Anderson insulators with disorder-localized single particle states \cite{imry_book2002} and very recently in Kagome 
metal CeRhSn \cite{kimura_npjquantmat2025}. Recent transport measurements on the Kagome metal Ni$_{3}$In revealed similar NFL signatures, which were attributed to be arising from a flat band induced localization of the itinerant fermions \cite{checkelsky_natphys2024,si_natcom2024}.   
\begin{figure}
\begin{center}
\includegraphics[height=15.5cm,width=8.5cm,angle=0]{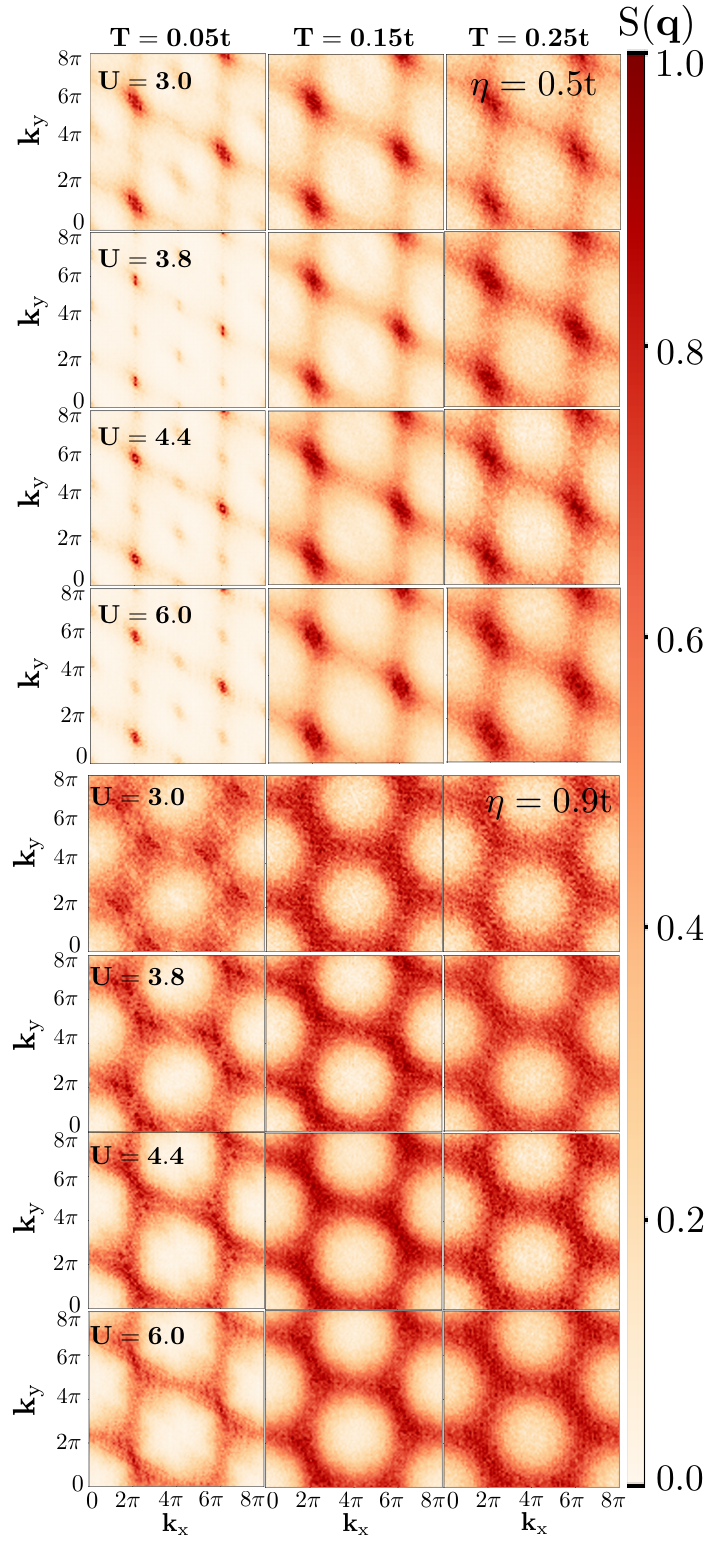}
\caption{Thermal evolution of the static magnetic structure factor ($S({\bf q})$) at $\eta=0.5t$ (intermediate strain) and $\eta=0.9t$ (large strain) 
for selected interactions, showing the progressive loss of magnetic correlations due to the thermal fluctuations induced disordering of the 
local magnetic moments. For each choice of $\eta$, temperature varies along the row while $U/t$ varies along the column.}
\label{fig10}
\end{center}
\end{figure}

\textit{Crossover to bad metal:} The breakdown of the Fermi liquid description is further analyzed based on the optical transport signatures viz. 
optical conductivity ($\sigma(\omega)$). The thermal evolution of the same is presented in Fig.\ref{fig9} across selected $U-\eta$ cross sections. 
The low temperature regime exhibits the suppression of the single particle transport at low frequencies and a DDP with a variable scaling 
exponent $\gamma$ such that, $\sigma(\omega) \propto 1/\omega^{\gamma}$ with $\gamma \neq 2$. Thermal fluctuations of the local moments
lead to the crossover of the NFL to the bad metal phase.  

The intermediate frequency optical spectra is characterized by a polaronic peak arising due to the strong correlation between the bosonic 
background and the itinerant fermions. Thermal fluctuations induced crossover of the NFL to the bad metal phase is quantified in terms of 
the melting of the polaronic peak, which defines the thermal scale $T_{IRM}(\eta)$ corresponding to the Ioffe-Regel-Mott temperature \cite{han_rmp2003,takagi_philmag2004}. The NFL-bad metal crossover is quantified in terms of the following criteria: (i) $A(T) \lesssim A(T_{0})/3$ and (ii) $\Gamma_{norm}=\Gamma/\omega_{peak} \ge 1$, where,  $A(T)$, $\Gamma$ and $\omega_{peak}$ correspond to the amplitude ($A(T_{0})$ is the amplitude at $T=0.01t$), the full width at half maxima (FWHM) and the polaronic peak frequency, respectively. These 
quantifiers are determined by fitting the intermediate frequency optical spectra to a Lorenzian, $g(x) = A(T)/(1+((x-\omega_{peak})/\Gamma)^{2})$ \cite{shashi_reentrant2025}. Based on these criteria, one obtains for $U=3.0t$, $\eta=0.6t$ the $T_{IRM} \sim 0.1t$ while for $\eta=0.9t$, $T_{IRM} \sim 0.15t$. 
\begin{figure*}
\begin{center}
\includegraphics[height=6.0cm,width=16.5cm,angle=0]{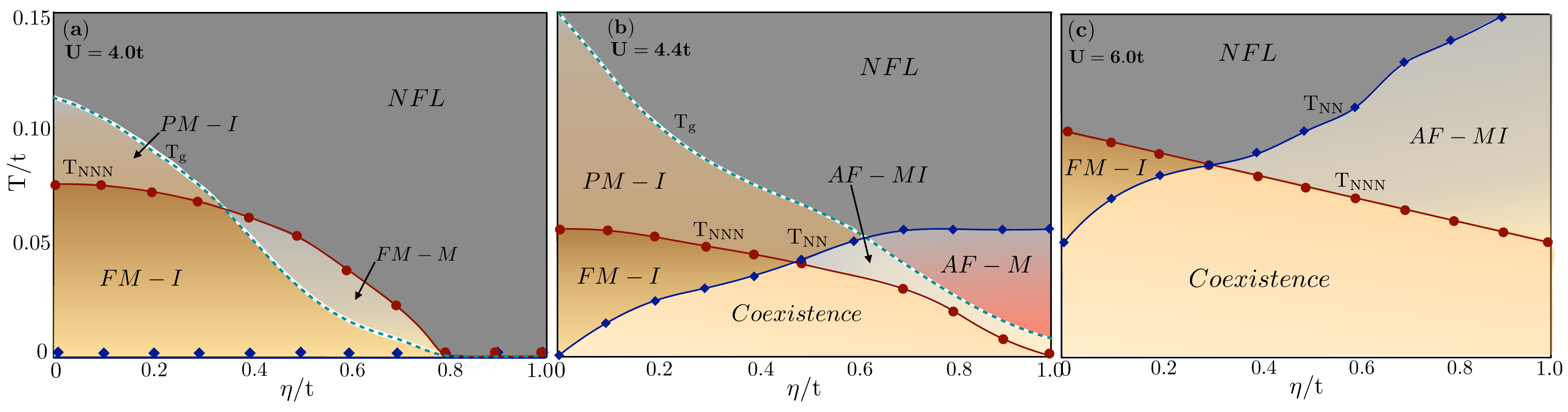}
\caption{Thermal phase diagram in the $\eta-T$ plane at selected interactions representing various thermodynamic phases. The 
thermal transition scales $T_{NN}$ and $T_{NNN}$ correspond to the thermal fluctuations induced loss of the NN and the NNN magnetic correlations, while $T_{g}$ represents the crossover thermal scale between the gapped and gapless phases (see text).}
\label{fig11}
\end{center}
\end{figure*}

The applied strain stabilizes the NFL phase against thermal fluctuations. The observation is generic across the $U-\eta$ plane. Note that the NFL to bad metal crossover has been reported in a wide class of materials such as, 2D TMD \cite{trivedi_prl2014,mak_nature2021,pasupathy_nature2021}, organic charge transfer salts \cite{fratini_natcom2021,kanoda_prl2020}, rare-earth nickelates \cite{stemmer_sciadv2021} and more recently Kagome metals \cite{tsirlin_prb2021}; with doping, temperature and electrical gating serving as the suitable control parameters.  

\textit{Static magnetic structure factor:}
In order to understand the fate of the magnetic correlations across the $U-\eta-T$ space we track the corresponding $S({\bf q})$ as shown in 
Fig.\ref{fig10} at the selected strain of $\eta=0.5t$ and $0.9t$. At $\eta=0.5t$ the magnetic correlations across the $U \gtrsim 3t$ regime 
(along the column) is largely the same at low temperatures with  $S({\bf q})$ exhibiting coexisting (quasi) long range magnetic order dominated 
by FM correlations with a finite but subdominant AF component. Thermal fluctuations (along the row) broaden the $S({\bf q})$ peaks allowing 
them to overlap, with the high temperature phase merely mapping out the non interacting Brillouin zone of the underlying lattice. 

At $\eta=0.9t$,  the system at low temperatures undergo cross over from one magnetic order to another as the interaction is tuned. 
The $\eta=0.9t, U=3t, T=0.05t$ cross section shows the reminiscent of the precursor Coulomb phase. Increasing interaction leads to the 
Coulomb phase. Thermal fluctuation results in the disordering of the (quasi) long range magnetic correlations, broadens the $S({\bf q})$ peaks 
and eventually maps out the Brillouin zone akin to that of the Kagome lattice. The magnetic correlations remain largely immune to the 
interactions for $U \ge 4.4t$ in the large strain limit, while at $U=3.8t$, thermal fluctuations destroy the precursor Coulomb phase to give 
rise to the magnetically disordered state. 
 
\textit{Thermal phase diagram:}
We next elucidate the thermal phases and the associated scales across the $U-\eta$ plane and distill the corresponding phase diagram in 
terms of the: (i) temperature dependent NN and the NNN magnetic correlations and (ii) thermal evolution of the single particle DOS. The 
crossover between the gapped and gapless phases is typified by the single particle DOS in terms of the thermal scale $T_{g}$.  We select 
three particular $U$'s viz. $U=4.0t$, $U=4.4t$ and $U=6.0t$, representative of the three different low temperature regimes and present the corresponding finite temperature $\eta-T$ phase diagram in Fig.\ref{fig11}.

At $U=4.0t$ (Fig.\ref{fig11}(a)), the thermal transition scale $T_{NNN}$ demarcates the FM correlated phase from the PM phases. The AF 
correlations are however strongly suppressed across the regime of strain. Over the regime $0 < \eta \lesssim 0.8t$ the low temperature phase 
is FM-I. Thermal fluctuations though destroys the magnetic correlations at $T_{NNN}$, don't close the single particle spectral gap at the Fermi level. This brings forth a magnetically disordered gapped insulating phase PM-I over the temperature regime $0.07t < T \lesssim 0.11t$, demarcated by the thermal scale $T_{g}$. We understand this phase as follows: while thermal fluctuations destroy the angular correlations between the local moments ${\bf m}_{i}$ and randomizes them, the strong coupling ensures a large $\vert {\bf m}_{i}\vert$ which couldn't get suppressed over this narrow temperature regime. The PM-I phase can therefore be envisaged as a regime of randomly oriented large local moments which though fails to order magnetically, don't close the excitation gap at the Fermi level. The gap closure takes place for $T \ge T_{g}$ leading to a NFL metal with short range magnetic correlations between the suppressed local magnetic moments. Over a narrow regime of strain $0.6t < \eta \lesssim 0.8t$, a finite temperature FM-M phase obtains, quantified by $T_{g} < T \lesssim T_{NNN}$, corresponding to a FM correlated gapless state. 

The $U=4.4t$ (Fig.\ref{fig11}(b)) cross section comprises of a FM-I phase close to the Lieb limit, over the temperature regime $0 < T \lesssim 0.04t$, with a dominant $T_{NNN}$ scale. Thermal fluctuations promote the AF correlations while simultaneously suppressing the FM order, 
giving rise to a finite temperature coexistence phase with competing $T_{NNN}$ and $T_{NN}$ scales. While $T_{NNN}$ and the corresponding FM correlations are strongly suppressed at $\eta \sim 0.8t$, the AF correlations are dominant over this regime and $T_{NN}$ saturates to a finite value at large strain. This finite temperature AF phase can further be demarcated into AF-M and AF-MI regimes based on $T_{g}$ determined 
from the single particle spectra. Further, a significant finite temperature PM-I regime is mapped out by $T_{g}$ at weak and intermediate strain, quantified by large randomly oriented local moments with a finite single particle spectral gap at the Fermi level. The $T > T_{g}$ regime essentially corresponds to a NFL metal. 
 
The low temperature regime at the strong coupling with $U=6t$ (Fig.\ref{fig11}(c)) is characterized by a FM-I for $0 < \eta \lesssim 0.9t$ and a AF-MI for $\eta \ge 0.9t$. Thermal fluctuations significantly enhances the AF correlations leading to a large $T_{NN}$ scale that increases with $\eta$,  at the same time weakly suppressing the $T_{NNN}$ scale corresponding to the FM correlations. A large part of the $\eta-T$ plane therefore hosts a coexistence phase with competing FM and AF correlations. The weak strain regime continues to be a FM-I at finite temperatures while the large strain regime hosts a finite temperature AF-MI phase. The single particle spectra is gapped across the $\eta-T$ plane corresponding to an insulator. The regime of $T \ge T_{NNN}$ and $T \ge T_{NN}$ is a NFL metal with finite spectral weight at the Fermi level in the single particle spectra. Note that at higher temperature a thermal crossover scale $T_{IRM}$ quantifies the crossover from the NFL 
to the bad metal phase independent of $U/t$ and $\eta/t$. 

\section{Discussion and conclusion}

Fig.\ref{fig6} and Fig.\ref{fig11} constitutes the primary results of this work wherein we have mapped out the low temperature phases and 
thermal scales in the $U-\eta$-plane. Our results on the thermodynamic, spectroscopic and transport properties have brought forth the applied strain as a suitable tuning parameter for the Lieb/Kagome interconversion; a cleaner alternative to disorder and chemical doping. The 
straintronics protocol discussed in this paper is generic and should be applicable to other line-graph lattices as well. 

The calculations are carried out using the SPA Monte Carlo simulation based on adiabatic approximation wherein the fast moving fermions are subjected to the spatially inhomogeneous background of slow thermal bosons. SPA  can suitably capture the temperature regime of $T > T_{FL}$. This is particularly important in case of geometrically frustrated systems such as, Kagome metal whose low temperature physics evades the standard numerical tools owing to severe fermion sign problem and system size restrictions. 

The NFL metal observed both at the low and finite temperatures is characterized by the deviation of the transport signatures from the Fermi 
liquid description. Of particular importance is the strain dependent variable scaling exponents of the low temperature longitudinal resistivity ($\rho_{xx}(T) \propto T^{\alpha}$) and low frequency optical conductivity ($\sigma(\omega) \propto 1/\omega^{\gamma}$). Similar observations of variable transport scaling exponents have been made in case of heavy fermion systems \cite{assmus_prl1998,steglich_prl2000,lonzarich_jpcm1996,ramazashvili_jpcm2001}, disorder induced phase transition \cite{imry_book2002,dagotto_prl2017} and more recently in Kagome metals \cite{kimura_npjquantmat2025}, thus establishing 
the fermionic localization by a fluctuating random background of slow bosons as a plausible explanation for the breakdown of the Fermi 
liquid description. 

It must be noted that the transient localization observed in the weak coupling regime of the system discussed in the preceding sections is a natural outcome of fermions interacting with the thermal bosons \cite{ciuchi_scipost2021,ciuchi_arxiv2024,kivelson_pnas2023,fratini_prb2023}. This PM-FI regime is weakly localized and is therefore prone to system size effects. In 2D, a finite sized system can host a weakly localized phase at weak coupling when $L < \xi$ (where, $\xi$ is the localization length) \cite{anderson1979}. Across the Lieb/Kagome interconversion, the (weak) transiently localized regime constitutes a stable low temperature phase in the $U-\eta-T$ space, for system sizes accessible within reasonable computation cost \cite{shashi_reentrant2025}. The key drawback of the adiabatic approximation of SPA is its failure to capture the $T<T_{FL}$ regime, where the translation invariance is restored and the system behaves as a conventional Fermi liquid,  as experimentally observed in 
several classes of strongly correlated quantum materials \cite{fratini_natcom2021,kanoda_prl2020,mak_nature2021,pasupathy_nature2021,tsirlin_prb2021}. 

One of the suitable platforms to realize the strain controlled metal-insulator transition discussed in this work is MOF. Phenomena such as, charge transfer, gate tuning and strain engineering are shown to reveal strong electronic correlation effects in MOF \cite{schiffrin_natcom2024,clerac_natcom2022,schiffrin_advfuncmat2021,medhekar_npjcompmat2022,bredas_materhor2022,hasimoto_sciadv2021,heine_chemsocrev2020,feng_chemsocrev2021}. Unconventional superconductivity \cite{hasimoto_sciadv2021}, exchange correlation induced ferromagnetism \cite{medhekar_npjcompmat2022} and recently gate controlled Mott insulator-metal transitions \cite{schiffrin_natcom2024} are reported in MOFs, making them a suitable and promising platform to realize straintronics across the Lieb/Kagome interconversion.   

\section{Acknowledgements} 
MK would like to acknowledge  the use of the high performance computing facility (AQUA) at the Indian 
Institute of Technology, Madras, India. MK acknowledges the support from the Department of Science and Technology, Govt. of India 
through the grant CRG/2023/002593.

\bibliography{magtrans.bib}
\end{document}